\pdfoutput=1

\documentclass[11pt]{article}

\usepackage{acl}

\usepackage{times}
\usepackage{latexsym}

\usepackage[T1]{fontenc}

\usepackage[utf8]{inputenc}

\usepackage{microtype}

\usepackage{inconsolata}

\usepackage{graphicx}

\usepackage{graphicx}
\usepackage{booktabs}
\usepackage{amsmath}
\usepackage{longtable}
\usepackage{hyperref}
\usepackage{float}
\usepackage{caption}
\usepackage{subcaption}
\title{Exploring AI-Enabled Test Practice, Affect, and Test Outcomes in Language Assessment}
\author{
{Jill Burstein}\thanks{Authors are listed alphabetically to reflect equal contributions.} \\
\textbf{Ramsey Cardwell} \\
\textbf{Ping-Lin Chuang} \\
\textbf{Allison Michalowski} \\
\textbf{Steven Nydick} \\\
Duolingo \\
\texttt{\{jill, ramsey, pinglin, allison.michalowski, steven.nydick\}@duolingo.com}
}
\date{}

\begin{document}

\maketitle

\begin{abstract}
Practice tests for high-stakes assessment are intended to build test familiarity, and reduce construct-irrelevant variance which can interfere with valid score interpretation. Generative AI-driven,  automated item generation (AIG) scales the creation of large item banks and multiple practice tests, enabling repeated practice opportunities. We conducted a large-scale observational study (N = 25,969) using the Duolingo English Test (DET)---a digital, high-stakes, computer-adaptive English language proficiency test to examine how increased access to repeated test practice relates to official DETscores, test-taker affect (e.g., confidence), and score-sharing for university admissions. To our knowledge, this is the first large-scale study exploring the use of AIG-enabled practice tests in high-stakes language assessment. Results showed that taking 1-3 practice tests was associated with better performance (scores), positive affect (e.g., confidence) toward the official DET, and increased likelihood of sharing scores for university admissions for those who also expressed positive affect. Taking more than 3 practice tests was related to lower performance, potentially reflecting \textit{washback} -- i.e., using the practice test for purposes other than test familiarity, such as language learning or developing test-taking strategies.  
Findings can inform best practices regarding AI-supported test readiness. Study findings also raise new questions about test-taker preparation behaviors and relationships to test-taker performance, affect, and behaviorial outcomes. 
\end{abstract}
\section{Introduction}
For millions of international test takers, scores on high-stakes English language proficiency (ELP) assessments can profoundly impact their educational and professional goals. As a result, they engage in various  test preparation strategies. For example,  practice tests aim to build familiarity for a specific test; reading books and articles can improve English language reading skills; and, deliberate engagement in  conversations with peers and instructors can strengthen English language speaking and listening skills.  

This paper focuses on \textit{practice tests}. Practice tests aim to build test familiarity to reduce test-design-related \textit{construct-irrelevant variance} (CIV). CIV is associated with the introduction of factors unrelated to the skills a test is intended to measure (the \textit{target construct}) \cite{Messick1982, Powers1985}. For instance, CIV can stem from unfamiliar technical features (e.g., \textit{drag-and-drop}), lack of familiarity with the device required for taking a test (e.g., test requirements to use a laptop for test takers who have limited laptop experience (\citet{konewe}), or anxiety triggered by an unfamiliar format \citep{WinkeLim2017}.

Conventional practice tests, often developed by testing organizations,  aim to reduce CIV.  However, they typically contain a limited number of fixed forms, restricting opportunities for repeated test practice. Modern generative AI-powered automated item generation (henceforth, AIG) alleviates this constraint by enabling the creation of large item pools for digital practice tests. As a result, practice test generation can be scaled to support repeated practice test opportunities for test takers.

The Duolingo English Test (DET)is a digital, AI-driven, high-stakes, computer-adaptive ELP assessment used for international student university admissions.  The DET is taken by hundreds of thousands of test takers each year. 

To help test takers become familiar with the test, the DET offers a \textit{free} practice test that simulates the official DET. As such, the practice test provides exposure to the DET task types, mirroring the official test in both appearance and administration order. It also provides an estimated score range, giving test takers a sense of how they are likely to perform on the official test. Like the official DET, the practice test is also computer-adaptive, but drawing from a separate item pool than the official test. The large practice-test item pool, enabled by AIG, is used to dynamically generate versions of the practice test with different item sets, offering test takers repeated opportunities for practice \citep{naismith25}.\footnote{The practice test items are created using the same AIG methods as the official DET.}\footnote{Successive versions of OpenAI's GPT models were used to develop the practice test, reflecting generative AI advances.} 

The study presented in this paper examines how access to repeated test practice (i.e., the number of tests taken)---enabled by AIG--- relates to test-takers' official DET scores, test-taker affect (e.g., confidence), and test-takers' decision to share their official DET scores for university admissions.

\section{Background}
Language assessment research has examined various aspects of test preparation, including test-taker preparation preferences \citep{OSullivan2021}, the relationship between preparation and affect (such as anxiety) \citep{ChangRead2008, PowersAlderman1983, WinkeLim2017}, and the link between preparation and test performance \citep{green2007washback,knoch2020drawing, Liu2014, Powers1985, xie2013does}.  These studies suggest that test preparation can reduce anxiety \citep{ChangRead2008, PowersAlderman1983}, increase confidence \citep{PowersAlderman1983}, and improve test scores \citep{green2007washback,knoch2020drawing, xie2013does}. \citet{knoch2020drawing} investigated repeat test takers, showing how they changed their test preparation strategies over time to try to improve their test score. \citet{xie2013does} demonstrated how test takers use test preparation to develop strategies for score improvement. \citet{green2007washback} examined the comparative impact of test preparation courses for a high-stakes language assessment. These three studies highlight \textit{washback effect} with regard to test preparation, whereby a test influences language teaching and learning \cite{messick1996validity}. 

Automated item generation research related to assessment and instruction is extensive, but much predates modern generative AI.  For example, \citet{MitkovHaKaramanis2006} showed that NLP-assisted item generation with human review can be more time-efficient than manual creation. \citet{HeilmanSmith2010} proposed a framework for automatically generating and evaluating questions from text, demonstrating the feasibility of transforming declarative sentences into fact-based questions. Similarly, \citet{MadnaniEtAl2016} discussed the Language Muse system, which used NLP to generate reading comprehension exercises for U.S. middle school texts for English learners. More recent research has shifted toward evaluating item quality and comparing system performance using large language models. For instance, \citet{laverghetta2023generating} investigated GPT-4 for test item generation, demonstrating its potential to create psychometrically valid items.\footnote{Also see \citet{flor2025question} for a comprehension discussion of automated item generation.} 

AIG is now integrated into the development of digital, high-stakes language assessments. Specific to this paper, the official DET and its practice test are dynamically assembled using AIG-created item banks with human review \citep{attali2022interactive}. After generating items with prompts used to fine-tune the AIG, human experts conduct a review. To ensure item quality and appropriateness, a multistage process for human review is implemented. This process begins with automated checks for linguistic accuracy and social appropriateness, followed by human expert review focused on copyediting, fact-checking, and identifying potential fairness and bias issues that could disadvantage certain test-taker groups \citep{church2025guidelines}.  

An internally-developed review platform is used to coordinate item reviews, track reviewer performance, and ensure inter-rater consistency. The final items are used to automatically create the DET practice and \textit{official} DET tests. 

 As mentioned earlier, prior research about test preparation for high-stakes assessment has studied test-taker preferences, and established links between test preparation, test-taker affect, and performance outcomes. However, we are unaware of research examining how test takers' access to repeated practice tests---now enabled by AIG---relates to these factors. This likely stems from the limited scalability of conventional practice tests, which rely on human test developers who cannot generate test items at the same scale as AIG. \citet{he2024scoping} conducted an extensive literature review, including 66 studies about research for second language test preparation. No themes emerged demonstrating research that examined technology or AI to enhance test preparation. 
\enlargethispage{-1.0cm}
\section {The Study} 
 This \textit{observational study}  examined how access to repeated practice test opportunities---enabled by AIG---related to test takers' official DET performance, test-taker affect, and test-taker decisions to share their official DET scores for university admissions. The study addressed the research question: What are the \textit{observed} relationships between the number of practice tests taken and test-takers' official DET performance, test-taker affect, and test-taker score sharing decisions?

\subsection {Methods}
\subsubsection{Survey instrument}
\begin{figure}
    \centering
    \includegraphics[width=0.5\textwidth]{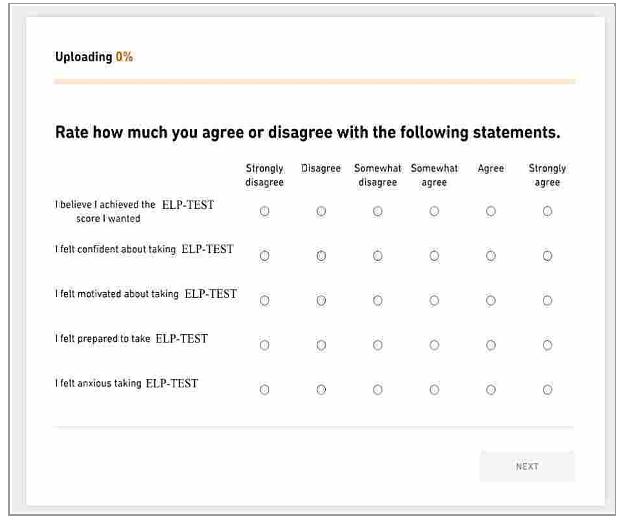} 
    \caption{Post–DET Affective Perceptions Survey}
    \label{fig:survey}
\end{figure}

To measure test takers’ affect, we developed a brief survey instrument (henceforth, \textit{survey}) that elicited perceptions of \textit{achievement, confidence, motivation, preparedness}, and \textit{anxiety} in relation to the official DET. The survey items reflect affective factors commonly used in prior research on assessment \citep[e.g.,][]{WinkeLim2017} and instructional contexts \citep[e.g.,][]{ling2021writing}. We acknowledge that typical affective surveys include more items per construct. However, because the DET is an operational, high-stakes assessment, there are required constraints: we had to limit the number of post-test, \textit{offboarding}\footnote{Offboarding takes place once the test is completed. Test takers are asked questions related to, e.g., demographics and their target score.} questions to avoid overburdening test takers. Consequently, the survey consisted of five items, each rated on a six-point Likert-style scale. The survey was presented to all test takers as shown in \textbf{Figure~\ref{fig:survey}}.
 
\subsubsection{Data Collection}  
The survey was administered during September 2023. Upon completion of the DET, test takers were presented with the survey during the DET offboarding process.  

Of the original 32,599 test-taker participants (henceforth, test takers) who took the survey, responses were retained from 25,969 test-takers for the analysis. Responses were retained only for participants who: (1) responded to all survey items; (2) were taking the official DET for the first time\footnote{Prior testing may have provided additional practice, complicating the analysis.}; (3) received an official DET score that was validated by human proctors; and,  (4) had taken the practice tests within 60 days prior to taking the official DET.  

\textbf{Table~\ref{tab:correlations}} shows the Spearman rank-order correlations between the survey items. The pairwise correlations between \textit{I believed I achieved the DET score I wanted} (Achieved), \textit{I felt confident about taking the DET} (Confident), \textit{I felt motivated about taking the DET} (Motivated), and \textit{I felt prepared to take the DET} (Prepared) are moderately high. This suggests that these positive affective statements may be related to a similar construct. By contrast, \textit{I felt anxious taking the DET} (Anxious) is effectively uncorrelated with the other items.
\begin{table}
  \centering
  \begin{tabular}{lccccc}
    \hline
    & \textbf{ACH} & \textbf{CON} & \textbf{MOT} & \textbf{PREP} & \textbf{ANX} \\
    \hline
    ACH & 1.00 & 0.74 & 0.59 & 0.67 & 0.04 \\
    CON & 0.74 & 1.00 & 0.68 & 0.72 & -0.03 \\
    MOT & 0.59 & 0.68 & 1.00 & 0.64 & 0.07 \\
    PREP  & 0.67 & 0.72 & 0.64 & 1.00 & 0.06 \\
    ANX   & 0.04 & -0.03 & 0.07 & 0.06 & 1.00 \\
    \hline
  \end{tabular}
  \label{tab:correlations}
  \caption{Spearman Correlations Between Responses to Survey Items; ACH=Achieved; CON=Confident; MOT=Motivated; PREP=Prepared; ANX=Anxious}
  \label{tab:correlations}
\end{table}

\subsubsection {Participant Demographics}
Test taker demographic information is collected from test takers during the official DET's offboarding process. Offboarding items ask test takers about their \textit{gender}, \textit{age}, \textit{testing intent} (i.e., obtaining an undergraduate or graduate degree), and \textit{first language}.\footnote{One hundred unique languages were reported by at least five participants.} 
\textbf{Table~\ref{tab:demographics}} shows the self-reported, test-taker demographics, also comparing the participant sample to the DET test-taker population \citep{naismith25}. The sample includes all demographic subgroups from the DET population, though with some variation in proportions. This may be because the study included only first-time test takers, while the DET test-taker population includes both first-time and repeat test takers.

\begin{table}
  \centering
  \begin{tabular}{lcc}
    \hline
    \textbf{Demographic} & \textbf{TTs (\%)} & \textbf{DET(\%)} \\
    \hline
    \textit{Gender} & & \\
    Female & 44.0 & 47.6 \\
    Male & 55.9 & 52.3 \\
    \hline
    \textit{Age Group} & & \\
    16--20 years & 19.0 & 32.7 \\
    21--25 years & 36.6 & 34.1 \\
    26--30 years & 18.8 & 14.8 \\
    \hline
    \textit{Testing Intent} & & \\
    Undergraduate & 43.0 & 47.1 \\
    Graduate & 43.7 & 37.0 \\
    \hline
    \textit{First Language} & & \\
    English & 13.7 & 9.5 \\
    Mandarin & 10.8 & 17.8 \\
    Telugu & 10.3 & 5.8 \\
    Spanish & 8.8 & 10.0 \\
    Arabic & 5.9 & 5.1 \\
    \hline
  \end{tabular}
  \caption{Test-Taker Demographics; TTs=Test takers from this study; DET=DET population}
  \label{tab:demographics}
\end{table}
\subsection{Analyses}
This section discusses relationships that emerged between test takers' DET practice test engagement (i.e, \textit{number of practice tests taken}), and their official DET scores, their affect (as self-reported in the survey), and their score-sharing decisions.\footnote{We used test takers' unique, official DET IDs to link to their practice test activity and score report sharing.} 

\textbf{Table~\ref{tab:summary_practice_tests}} shows official DET scores by number of practice tests taken.  Test takers were grouped into six bins (\textit{count groups)} by number of practice tests completed (0, 1, 2–3, 4–6, 7+). We chose these categories to distinguish between 0, 1, and multiple practice test-taking sessions.  Multiple practice test counts were grouped to balance the bin sample sizes. 

Table~\ref{tab:summary_practice_tests} suggests a relationship between practice tests taken and official DET scores. For each practice test count group, we included 95\% confidence intervals of the mean test score. The highest average scores were observed among those who took 1–3 practice tests (in \textbf{bold rows}). Confidence intervals of the mean test score for these rows did not overlap with those for 0, or 4 or more practice tests, showing significant differences.  Those who took 0, or 4 or more practice tests scored slightly lower.\footnote{Average scores across all groups hovered around the B2 CEFR level—a benchmark for independent language users and a common minimum for admission to English-medium universities \cite{CEFR2020}. However, it is important to note that where the test taker sits in the B2 CEFR range (lower vs. higher in the range) can impact their acceptance to a university.} 

The finding that scores do not continue to increase with 4 or more practice tests aligns with expectations: practice tests are intended to build test familiarity, which on its own, should not facilitate large jumps in language proficiency. 
\begin{table}
  \centering
  \setlength{\tabcolsep}{4pt}  
  \begin{tabular}{@{}lrccc@{}}
    \toprule
    \textbf{\# of PT} & \textbf{N} & \textbf{\%} & \textbf{M} & \textbf{CI 95\%}\\
    \midrule
    0     & 4,742  & 18.3& 108.5& [107.8, 109.2]\\
    \textbf{1}     & \textbf{6,128}  & \textbf{23.6}& \textbf{112.4}& \textbf{[111.8, 113.0]}\\
    \textbf{2--3}  & \textbf{6,469}  & \textbf{24.9}& \textbf{112.3}& \textbf{[111.8, 112.8]}\\
    4--6  & 4,142  & 16.0& 111.1& [110.5, 111.7]\\
    7+& 4,488& 17.3& 108.6& [108.1, 109.1]\\
    Total & 25,969&&&\\
    \hline
  \end{tabular}
  \caption{Mean (M) Overall DET Score by Number of Practice Tests Taken (\# of PT)} 
  \label{tab:summary_practice_tests}
\end{table}

\textbf{Table~\ref{tab:mean_ELP_scores}} illustrates the relationship between number of practice tests taken, test-taker affect, and test takers' official DET score. As no clear differences emerged across the original Likert-scale categories  ({Figure~\ref{fig:survey}), the six Likert-scale categories were collapsed into two. \textit{Agree} contained: Strongly Agree, Agree, and Somewhat Agree. and \textit{Disagree} contained: Strongly Disagree, Disagree, and Somewhat Disagree. 

We included 95\% confidence intervals of the difference between the mean scores for those who Agree and Disagree.\footnote{The Disagree mean score was subtracted from the Agree mean score.} Rows in \textbf{bold} indicate that the confidence interval did not include 0, showing significant differences.
{Table~\ref{tab:mean_ELP_scores} consistently shows that among test takers who took 0–3 practice tests, those who Agreed with positively-oriented items (Achieved, Confident, Motivated, Prepared) performed significantly better on the official DET than those who Disagreed. For those who Agreed they were Motivated and Prepared, better performance was also observed for 7+, and 4-6 and 7+ groupings, respectively. 

Test takers who took 0 or 1 practice test showed a significant score difference between those who Agreed and Disagreed across all positive statements.  
As well, test takers who practiced 2-3 times also showed significant differences between those who Agreed and Disagreed with the positive statements. This finding suggests that for some test takers, access to repeated test practice was related to positive affect and higher test scores. 

Across the large proportion of test takers who indicated they felt Anxious (70.8\%-75.3\%), there was no signficant relationship found based on the number of practice tests taken. A possible explanation is the high-stakes nature of the DET. In recent work in classroom settings, \citet{dehobeyond} found relationships between test anxiety and demographic factors. This is something that could be explored in future research. 

\textbf{Table~\ref{tab:sharedscores}} indicates a relationship between number of practice tests taken, likelihood of score sharing for university admissions, and test-taker affect. 

We used 95\% confidence intervals for the share rates (proportions) of those who Agreed or Disagreed with each of the statements. Rows  in \textbf{bold} indicate that the corresponding Agree and Disagree confidence intervals did not overlap, which showed significant differences.  
Test takers who took 0, 1, or 2-3 practice tests and Agreed with the Achieved, Confident, and Prepared statements had non-overlapping confidence intervals with test takers who took 0, 1, or 2-3 practice tests and Disagreed with those statements. For those who Agreed with the Motivated statement, only those who took 2-3 practice tests had share rate confidence intervals that did not overlap with the corresponding confidence intervals with those who Disagreed. Note that test takers were always more likely to share their scores if they Agreed with positive statements. 
 
As expected, further analysis showed that test takers who shared their scores tended to have higher mean scores. Scores typically aligned with a mid- to high B2 CEFR level. This is an expected outcome, as test takers are more likely to share scores that meet university requirements. Scores were highest among those who took 0–3 practice tests and Agreed with positive sentiment statements. For example, those who Agreed with the Achieved category had mean scores of 119.1, 120.9, and 119.0 for 0, 1, and 2–3 tests taken, respectively. This trend held across all positive sentiment categories. Scores declined slightly for those who took 4–6 tests (about 1 point lower) and more noticeably for those with 7+ tests (about 3 points lower). A similar pattern emerged for the Anxious category.


\section{Discussion}
Integrated into the DET pipeline, AIG generates large item pools. This scales the creation of 
DET practice tests, which increases test takers' access to repeated practice opportunities. To our knowledge, this is the first study to examine how AIG can contribute to increased practice opportunities and how, in turn, access to more practice is related to test-taker affect and outcomes. The study explored relationships between (1) practice test engagement and test score. (Table~\ref{tab:summary_practice_tests}), (2) test-taker affect and official DET scores (Table~\ref{tab:mean_ELP_scores}), and (3) affect and score-sharing decisions for university admissions (Table~\ref{tab:sharedscores}). 

Three key findings} emerged from the analysis to address our research question:  What are the \textit{observed} relationships between the number of practice tests taken,  and official DET performance, test-taker affect, and score-report sharing decisions?

\begin{table}[h] 
  \centering
  \begin{tabular}{lccccc}
    \hline
    \textbf{\#} & \multicolumn{2}{c}{\textbf{Agree}} & \multicolumn{2}{c}{\textbf{Disagree}} & \textbf{CI 95\%}\\
    & \textbf{\%}& \textbf{M}& \textbf{\%}& \textbf{M}& \\
    \hline
    \multicolumn{6}{c}{\textbf{Achieved}}\\
    \hline
   \textbf{ 0} & \textbf{85.7}& \textbf{109.5}& \textbf{14.3}& \textbf{102.2} & \textbf{[5.1, 9.5]}\\
    \textbf{1} & \textbf{82.8}& \textbf{113.8}& \textbf{17.2}& \textbf{105.6} & \textbf{[6.6, 9.9]}\\
    \textbf{2-3} & \textbf{84.3}& \textbf{112.8}& \textbf{15.7}& \textbf{109.1} & \textbf{[2.3, 5.3]}\\
    4-6 & 87.4& 111.3& 12.6& 109.7 & [-0.3, 3.5]\\
    7+& 91.0& 108.6& 9.0& 108.3& [-1.6, 2.2]\\
    \hline
    \multicolumn{6}{c}{\textbf{Confident}}\\
    \hline
    \textbf{0} & \textbf{85.4}& \textbf{109.9}& \textbf{14.6}& \textbf{100.5}& \textbf{[ 7.2, 11.6]}\\
    \textbf{1} & \textbf{82.8}& \textbf{114.0}& \textbf{17.2}& \textbf{104.6}& \textbf{[ 7.8, 11.1]}\\
    \textbf{2-3} & \textbf{84.4}& \textbf{113.2}& \textbf{15.6}& \textbf{107.4}& \textbf{[4.2, 7.2]}\\
    4-6 & 86.6& 111.3& 13.4& 109.6& [-0.2, 3.6]\\
    7+& 91.4& 108.7& 8.6& 108.3& [-1.6, 2.4]\\
    \hline
    \multicolumn{6}{c}{\textbf{Motivated}}\\
    \hline
    \textbf{0} & \textbf{90.9}& \textbf{109.1}& \textbf{9.1}& \textbf{102.1}& \textbf{[4.1, 9.9]}\\
   \textbf{ 1} & \textbf{89.8}& \textbf{113.0}& \textbf{10.2}& \textbf{107.5}& \textbf{[3.2, 7.7]}\\
    \textbf{2-3} & \textbf{91.7}& \textbf{112.6}& \textbf{8.3}& \textbf{108.8}& \textbf{[1.7, 5.8]}\\
    4-6 & 93.0& 111.2& 7.0& 110.1& [-1.6, 3.7]\\
   \textbf{ 7+}& \textbf{95.2}& \textbf{108.8}& \textbf{4.8}& \textbf{105.8}& \textbf{[0.4, 5.5]}\\
    \hline
    \multicolumn{6}{c}{\textbf{Prepared}}\\
    \hline
   \textbf{ 0} & \textbf{85.3}& \textbf{110.0}& \textbf{14.7}& \textbf{99.9}& \textbf{[ 7.8, 12.2]}\\
   \textbf{ 1} & \textbf{82.5}& \textbf{114.3}& \textbf{17.5}& \textbf{103.6}& \textbf{[ 9.1, 12.3]}\\
   \textbf{ 2-3} & \textbf{85.4}& \textbf{113.3}& \textbf{14.6}& \textbf{106.2}& \textbf{[5.5, 8.6]}\\
   \textbf{ 4-6} & \textbf{88.3}& \textbf{111.6}& \textbf{11.7}& \textbf{107.3}& \textbf{[2.3, 6.2]}\\
    \textbf{7+}& \textbf{92.7}& \textbf{108.9}& \textbf{7.3}& \textbf{105.4}& \textbf{[1.4, 5.5]}\\
    \hline
    \multicolumn{6}{c}{\textbf{Anxious}}\\
    \hline
    0 & 70.8& 107.6& 29.2& 110.6& [-4.6, -1.6]\\
    1 & 72.9& 112.2& 27.1& 113.1& [-2.2, 0.4]\\
    2-3 & 73.9& 112.3& 26.1& 112.1& [-0.9, 1.4]\\
    4-6 & 75.3& 111.2& 24.7& 110.6& [-0.7, 1.9]\\
    7+& 75.0& 108.5& 25.0& 108.9& [-1.5, 0.8]\\
    \hline
  \end{tabular}
  \caption{Mean (M) Overall DET Score by Practice Tests Taken (\#) and Affective Perceptions}
  \label{tab:mean_ELP_scores}
\end{table}

\textbf{First, repeated test practice was related to higher test scores to an extent. (Table~\ref{tab:summary_practice_tests})}.  Those who took 1, or 2-3 practice tests had comparatively higher scores than those who took 0, or more than 3.  As taking 2-3 practice tests was related to higher test scores, this suggests a potential benefit of access to repeated practice for some test takers.  These test takers may have come to the practice test with higher proficiency and were using the practice test for its intended purpose---i.e., test familiarity.

By contrast, taking more than 2-3 practice tests was associated with lower performance. This may be related to washback effect (mentioned earlier). 
Specifically, test takers may have used the practice test for reasons beyond test familiarity, such as building English language skills (i.e., positive washback that supports language learning), or test-taking strategies, such as trying to \textit{game} the test (i.e., negative washback that does not support language learning) \cite{knoch2020drawing,xie2013does}. In this scenario, test takers' repeated practice testing may be an example of \textit{wheel spinning}, where learners repeated attempts to master a skill are unsuccessful \cite{BeckGong2013,mu2020towards}.
\begin{table}[h] 
  \centering
  \begin{tabular}{lcccc}
    \hline
    \textbf{\#} & \multicolumn{2}{c}{\textbf{Agree}}& \multicolumn{2}{c}{\textbf{Disagree}}\\
 & \textbf{\%}& \textbf{CI 95\%}& \textbf{\%}&\textbf{CI 95\%}\\
    \hline
    \multicolumn{5}{c}{\textbf{Achieved}}\\
    \hline
   \textbf{ 0} & \textbf{41.9}&\textbf{[40.3, 43.4]}& \textbf{32.2}&\textbf{[28.6, 35.7]}\\
   \textbf{ 1} & \textbf{43.5}&\textbf{[42.1, 44.8]}& \textbf{31.9}&\textbf{[29.1, 34.7]}\\
    \textbf{2-3}& \textbf{43.4}&\textbf{[42.1, 44.7]}& \textbf{33.6}&\textbf{[30.7, 36.5]}\\
 4-6& 42.2& [40.6, 43.8]& 38.0&[33.9, 42.2]\\
 7+& 44.3& [42.8, 45.8]& 40.2&[35.5, 45.0]\\
    \hline
    \multicolumn{5}{c}{\textbf{Confident}}\\
    \hline
    \textbf{0} & \textbf{42.2}&\textbf{[40.7, 43.7]}& \textbf{30.3}&\textbf{[26.8, 33.7]}\\
    \textbf{1} & \textbf{43.6}&\textbf{[42.2, 44.9]}& \textbf{31.4}&\textbf{[28.6, 34.2]}\\
    \textbf{2-3}& \textbf{43.5}&[\textbf{42.2, 44.8]}& \textbf{32.8}&\textbf{[29.9, 35.7]}\\
 4-6& 42.1& [40.5, 43.7]& 38.8&[34.7, 42.8]\\
 7+& 44.0& [42.5, 45.5]& 43.3&[38.4, 48.2]\\
    \hline
    \multicolumn{5}{c}{\textbf{Motivated}}\\
    \hline
    0 & 41.0&[39.5, 42.4]& 35.5&[31.0, 40.0]\\
    1 & 41.9&[40.6, 43.2]& 37.9&[34.1, 41.8]\\
   \textbf{2-3}& \textbf{42.5}&\textbf{[41.2, 43.8]}& \textbf{34.7}&\textbf{[30.7, 38.7]}\\
 4-6& 41.7& [40.1, 43.2]& 41.2&[35.6, 46.9]\\
 7+& 44.2& [42.7, 45.6]& 39.6&[33.1, 46.1]\\
    \hline
    \multicolumn{5}{c}{\textbf{Prepared}}\\
    \hline
    \textbf{0} & \textbf{42.1}&\textbf{[40.6, 43.6]}& \textbf{31.1}&\textbf{[27.7, 34.5]}\\
    \textbf{1} & \textbf{43.5}&\textbf{[42.1, 44.9]}& \textbf{32.0}&\textbf{[29.2, 34.8]}\\
    \textbf{2-3}& \textbf{43.5}&\textbf{[42.2, 44.8]}& \textbf{32.3}&\textbf{[29.3, 35.3]}\\
 4-6& 42.1& [40.5, 43.7]& 38.0&[33.7, 42.3]\\
 7+& 44.3& [42.8, 45.8]& 38.9&[33.6, 44.2]\\
    \hline
    \multicolumn{5}{c}{\textbf{Anxious}}\\
    \hline
    0 & 39.7&[38.0, 41.3]& 42.4&[39.8, 45.0]\\
    1 & 41.1&[39.6, 42.5]& 42.6&[40.3, 45.0]\\
    2-3& 42.0&[40.6, 43.4]& 41.4&[39.1, 43.8]\\
 4-6& 41.4& [39.6, 43.1]& 42.5&[39.5, 45.6]\\
 7+& 43.2& [41.5, 44.8]& 46.3&[43.4, 49.2]\\
 \hline
  \end{tabular}
  \caption{Proportion of Test Takers who Shared Their DET Score by Number of Practice Tests Taken and Affective Perceptions} 
  \label{tab:sharedscores}
\end{table}




\textbf{Second, test takers who took more practice tests reported feeling more positively (Table~\ref{tab:mean_ELP_scores}).} Based on the number of practice tests taken, higher proportions of test takers reported positive affect toward the official DET regarding their beliefs that they achieved the score they wanted, and their confidence, motivation, and preparedness. As such, the 7+ group consistently had the highest proportion of test takers reporting positive affect. Reported feelings of anxiety were similar across the number of practice tests taken (Table~\ref{tab:mean_ELP_scores}).  While not surprising in a high-stakes  context, the finding is novel compared to prior work suggesting that test preparation could reduce anxiety \cite{ChangRead2008,PowersAlderman1983, WinkeLim2017}. However, previous work was conducted in no- or low-stakes experimental settings. 

Regarding DET performance, test takers who agreed with the positive statements had higher official DET scores, on average, than those who disagreed; this finding was significant (Table~\ref{tab:mean_ELP_scores}).} Those who took 1-3 practice tests had the highest scores, on average. Test scores trended lower after taking more than 3 practice tests. 

\textbf{Third, test takers who reported positive perceptions were more likely to share their official DET score report for university admissions (Table~\ref{tab:sharedscores}).}   This finding was consistent across the number of practice tests taken with comparatively higher proportions for those who Agreed than Disagreed with the positive survey items. Share rates were significantly higher for those who took 0-3 practice tests and Agreed with the Achieved, Confident and Prepared statements, and for those who took 2-3 practice tests and Agreed with the Motivation statement, as compared to those who Disagreed. Like other outcomes we investigated, Anxiety did not show significant differences in share rates by agreement status.
\section{Limitations}
This section notes \textit{two} study limitations.

\textit{ First},   as an observational study, our findings are \textbf{not causal}. Independent of practice test use, higher English proficiency may underlie positive perceptions, higher scores, and share rates.  

\textit{Second}, the number of survey items was necessarily limited to reduce the burden test takers after taking a high-stakes test. Given this real-world constraint, we prioritized items related to test-taker affect, and did not include an item eliciting information about alternative test strategies. As a result,  we lacked data on test takers' use of alternative preparation methods.  Related, we do not have information about what motivated test takers' repeated practice. As we continue with this research, we are exploring ways to address this limitation.

\section{Conclusions}
The DET’s practice test simulates the official DET. As a computer-adaptive test, the practice test aims to familiarize test takers with the official DET’s item types, its adaptive administration, and the official DET score scale (by providing an estimated test score range). Integrating AIG  into the test development pipeline enables scalable production of DET practice tests. This facilitates the creation of multiple practice test versions, offering test takers repeated opportunities to build test familiarity. 

The study analysis showed that test takers who took 1-3 practice tests tended to have higher official DET scores. Higher test scores were also related to positive affect (i.e., agreeing with the positive survey items). Higher share rates were also linked to positive affect. This \textit{may} be related to those test takers having higher underlying English proficiency. Therefore, test takers may have used the practice test for its intended purpose---test familiarization, whereby 1-3 practice test repetitions may have been sufficient. This also suggests that for some test takers---those who took 2-3 practice tests--- that \textit{limited} repeated practice may have provided extra needed support to sufficiently build their test familiarity.

By contrast, test takers who took more than 3 practice tests had lower performance, on average. It is possible that these test takers may have come to the test with lower proficiency. Their additional test practice may be related to washback, whereby test takers used the practice test for reasons besides building test familiarity (e.g., English language learning or building test-taking strategies). However, we lack data about test takers' preparation strategies, beyond the DET practice test, as well as test-taker goals for taking the practice test. Therefore, this limits interpretation. At the same time, it  raises interesting questions with regard to appropriate guidance about test preparation, especially with regard to mitigating negative washback effects, such as using test practice to develop test gaming strategies. 

AIG for high-stakes assessment is still in its early stages. 
The study examines how repeated practice—enabled by AIG—may relate to test-taker performance, affect, and behavioral outcomes (i.e., score sharing). It also raises important questions about test preparation practices when test-takers have access to repeated test practice. Our findings—and future research—could be useful in helping to inform best practices for AI-enhanced test readiness in high-stakes contexts.

\section*{Acknowledgements}
The authors would like to thank our Duolingo colleagues for helpful reviews of earlier versions of this paper: Audrey Kittredge,  Nitin Madnani, Ben Naismith, and Alina von Davier. Thanks to the anonymous reviewers for their feedback.

\bibliography{latex/acl_latex} 

\begin{thebibliography}{26}
\providecommand{\natexlab}[1]{#1}

\bibitem[{Attali et~al.(2022)Attali, Runge, LaFlair, Yancey, Goodwin, Park, and Von~Davier}]{attali2022interactive}
Yigal Attali, Andrew Runge, Geoffrey~T LaFlair, Kevin Yancey, Sarah Goodwin, Yena Park, and Alina~A Von~Davier. 2022.
\newblock The interactive reading task: Transformer-based automatic item generation.
\newblock \emph{Frontiers in Artificial Intelligence}, 5:1--13.

\bibitem[{Beck and Gong(2013)}]{BeckGong2013}
J.~E. Beck and Y.~Gong. 2013.
\newblock Wheel-spinning: Students who fail to master a skill.
\newblock In \emph{Artificial Intelligence in Education: 16th International Conference, AIED 2013}, pages 431--440. Springer.

\bibitem[{Chang and Read(2008)}]{ChangRead2008}
A.~C.~S. Chang and J.~Read. 2008.
\newblock Reducing listening test anxiety through various forms of listening support.
\newblock \emph{TESL-EJ}, 12(1).
\newblock N1.

\bibitem[{Church et~al.(2025)Church, Park, and Burstein}]{church2025guidelines}
Jacqueline Church, Yena Park, and Jill Burstein. 2025.
\newblock \href {https://go.duolingo.com/DETFairnessGuidelines} {Guidelines for fair test content: {The Duolingo English Test} example}.
\newblock Duolingo Research Report DRR-25-02, Duolingo.
\newblock 19 pages.

\bibitem[{{Council of Europe}(2020)}]{CEFR2020}
{Council of Europe}. 2020.
\newblock \href {https://www.coe.int/lang-cefr} {\emph{Common European Framework of Reference for Languages: Learning, Teaching, Assessment – Companion Volume}}.
\newblock Council of Europe Publishing.

\bibitem[{Deho et~al.(2025)Deho, Joksimovic, Vieira, and Baker}]{dehobeyond}
Oscar~Blessed Deho, Srecko Joksimovic, Maria Vieira, and Ryan Baker. 2025.
\newblock Beyond predictive accuracy: Fairness and bias in predicting test anxiety.
\newblock In \emph{Proceedings of the International Conference on Artificial Intelligence and Education}.

\bibitem[{Flor(2025)}]{flor2025question}
Michael Flor. 2025.
\newblock Question generation with large language models and generative ai.
\newblock In \emph{Automatic Question Generation}, pages 137--147. Springer.

\bibitem[{Green(2007)}]{green2007washback}
Anthony Green. 2007.
\newblock Washback to learning outcomes: A comparative study of ielts preparation and university pre-sessional language courses.
\newblock \emph{Assessment in Education}, 14(1):75--97.

\bibitem[{He et~al.(2024)He, S{\'e}n{\'e}cal, Stansfield, and Suvorov}]{he2024scoping}
Shanshan He, Anne-Marie S{\'e}n{\'e}cal, Laura Stansfield, and Ruslan Suvorov. 2024.
\newblock A scoping review of research on second language test preparation.
\newblock \emph{Language Testing}, 42(1):11--47.

\bibitem[{Heilman and Smith(2010)}]{HeilmanSmith2010}
Michael Heilman and Noah~A. Smith. 2010.
\newblock Question generation via overgenerating transformations and ranking.
\newblock Technical Report CMU-LTI-10-008, Language Technologies Institute, Carnegie Mellon University, Pittsburgh, PA.

\bibitem[{Knoch et~al.(2020)Knoch, Huisman, Elder, Kong, and McKenna}]{knoch2020drawing}
Ute Knoch, Annemiek Huisman, Cathie Elder, Xiaoxiao Kong, and Angela McKenna. 2020.
\newblock Drawing on repeat test takers to study test preparation practices and their links to score gains.
\newblock \emph{Language Testing}, 37(4):550--572.

\bibitem[{Kon{\'e} et~al.(2024)Kon{\'e}, Winke, and Gordon}]{konewe}
Kadidja Kon{\'e}, Paula Winke, and Matthew Gordon. 2024.
\newblock \href {https://osf.io/preprints/osf/tsbf5_v1} {“we would like to see ourselves in the test:” the experiences of francophone african english learners in high-stakes english proficiency testing}.

\bibitem[{Laverghetta~Jr and Licato(2023)}]{laverghetta2023generating}
Antonio Laverghetta~Jr and John Licato. 2023.
\newblock Generating better items for cognitive assessments using large language models.
\newblock In \emph{Proceedings of the 18th workshop on innovative use of NLP for building educational applications (BEA 2023)}, pages 414--428.

\bibitem[{Ling et~al.(2021)Ling, Elliot, Burstein, McCaffrey, MacArthur, and Holtzman}]{ling2021writing}
Guangming Ling, Norbert Elliot, Jill~C Burstein, Daniel~F McCaffrey, Charles~A MacArthur, and Steven Holtzman. 2021.
\newblock Writing motivation: A validation study of self-judgment and performance.
\newblock \emph{Assessing Writing}, 48:100509.

\bibitem[{Liu(2014)}]{Liu2014}
O.~L. Liu. 2014.
\newblock Investigating the relationship between test preparation and toefl ibt performance.
\newblock \emph{ETS Research Report Series}, (2):1--13.

\bibitem[{Madnani et~al.(2016)Madnani, Burstein, Sabatini, Biggers, and Andreyev}]{MadnaniEtAl2016}
Nitin Madnani, Jill Burstein, John Sabatini, Kristy Biggers, and Slava Andreyev. 2016.
\newblock Language muse\texttrademark: Automated linguistic activity generation for english language learners.
\newblock In \emph{Proceedings of the Annual Meeting of the Association for Computational Linguistics}, Berlin, Germany.

\bibitem[{Messick(1982)}]{Messick1982}
S.~Messick. 1982.
\newblock Issues of effectiveness and equity in the coaching controversy: Implications for educational and testing practice.
\newblock \emph{Educational Psychologist}, 17:67--91.

\bibitem[{Messick(1996)}]{messick1996validity}
Samuel Messick. 1996.
\newblock Validity and washback in language testing.
\newblock \emph{Language testing}, 13(3):241--256.

\bibitem[{Mitkov et~al.(2006)Mitkov, Ha, and Karamanis}]{MitkovHaKaramanis2006}
Ruslan Mitkov, Le~An Ha, and Nikiforos Karamanis. 2006.
\newblock A computer-aided environment for generating multiple-choice test items.
\newblock \emph{Natural Language Engineering}, 12(2):177--194.

\bibitem[{Mu et~al.(2020)Mu, Jetten, and Brunskill}]{mu2020towards}
Tong Mu, Andrea Jetten, and Emma Brunskill. 2020.
\newblock Towards suggesting actionable interventions for wheel-spinning students.
\newblock \emph{International Educational Data Mining Society}.

\bibitem[{Naismith et~al.(2025)Naismith, Cardwell, LaFlair, Nydick, and Kostromitina}]{naismith25}
B.~Naismith, R.~Cardwell, G.~LaFlair, S.~Nydick, and M.~Kostromitina. 2025.
\newblock \href {https://go.duolingo.com/dettechnicalmanual} {{Duolingo English Test}: Technical manual}.
\newblock Duolingo research report, Duolingo.

\bibitem[{O’Sullivan et~al.(2021)O’Sullivan, Dunn, and Berry}]{OSullivan2021}
B.~O’Sullivan, K.~Dunn, and V.~Berry. 2021.
\newblock Test preparation: An international comparison of test takers’ preferences.
\newblock \emph{Assessment in Education: Principles, Policy \& Practice}, 28(1):13--36.

\bibitem[{Powers(1985)}]{Powers1985}
D.~E. Powers. 1985.
\newblock Effects of test preparation on the validity of a graduate admissions test.
\newblock \emph{Applied Psychological Measurement}, 9(2):179--190.

\bibitem[{Powers and Alderman(1983)}]{PowersAlderman1983}
D.~E. Powers and D.~L. Alderman. 1983.
\newblock Effects of test familiarization on sat performance.
\newblock \emph{Journal of Educational Measurement}, 20(1):71--79.

\bibitem[{Winke and Lim(2017)}]{WinkeLim2017}
P.~Winke and H.~Lim. 2017.
\newblock \href {https://doi.org/10.1080/15434303.2017.1399396} {The effects of test preparation on second-language listening test performance}.
\newblock \emph{Language Assessment Quarterly}, 14(4):380--397.

\bibitem[{Xie(2013)}]{xie2013does}
Q.~Xie. 2013.
\newblock Does test preparation work? implications for score validity.
\newblock \emph{Language Assessment Quarterly}, 10(2):196--218.

\end{thebibliography}
\bibliographystyle{acl_natbib}

\end{document}